\documentclass[sigconf,table]{acmart}

\usepackage[utf8]{inputenc}
\usepackage{graphicx}
\usepackage{subfig}
\usepackage{xcolor}
\usepackage{amsmath,amssymb}
\usepackage{ifthen}
\usepackage{xspace}

\usepackage{algorithm}
\usepackage{algpseudocode}
\usepackage{color}
\usepackage{listings}
\usepackage{pifont}
\usepackage{multirow}
\usepackage{url}

\usepackage{listings}

\newcommand{\SYS}{\textsc{SecureStreams}\xspace}
\newcommand{\zmq}{\mbox{\textsc{ZeroMQ}}\xspace}
\newcommand{\rxl}{\mbox{\textsc{RxLua}}\xspace}
\newcommand{\luavm}{\mbox{\textsc{LuaVM}}\xspace}

\newboolean{showcomments}
\setboolean{showcomments}{true}
\ifthenelse{\boolean{showcomments}}
{ \newcommand{\mynote}[3]{
   \fbox{\bfseries\sffamily\scriptsize#1}
   {\small$\blacktriangleright$\textsf{\emph{\color{#3}{#2}}}$\blacktriangleleft$}}}
{ \newcommand{\mynote}[3]{}}

\definecolor{darkgreen}{rgb}{0.3,0.5,0.3}
\definecolor{darkblue}{rgb}{0.3,0.3,0.5}
\definecolor{darkred}{rgb}{0.5,0.3,0.3}

\lstdefinelanguage{LUA}{
  sensitive=true,
  keywordstyle=[1]{\color{darkblue}\bfseries},
  keywordstyle=[2]{\color{darkgreen}\bfseries},
  morekeywords=[1]{and,break,do,else,elseif,end,for,function,if,in,local,
    nil,not,or,repeat,return,then,until,while,require,alias},  morekeywords=[2]{},  otherkeywords={.,=,~,*,>,:},
  morestring=[b]",
  stringstyle={\color{darkred}\itshape},
  breaklines=true,
  breakatwhitespace=true,
  linewidth=\columnwidth,
  comment=[l]{--},
  escapeinside={(*@}{@*)}
}

\lstdefinelanguage{YAML}{
  sensitive=true,
  keywordstyle=[1]{\color{darkblue}\bfseries},
  keywordstyle=[2]{\color{darkgreen}\bfseries},
  morekeywords=[1]{image, entrypoint, environment, devices, hostname},  morekeywords=[2]{},  otherkeywords={.,=,~,*,>,:},
  morestring=[b]",
  breaklines=true,
  breakatwhitespace=true,
  linewidth=\columnwidth,
  comment=[l]{--},
  escapeinside={(*@}{@*)}
}

\setcopyright{rightsretained}

\setcopyright{acmcopyright}
\acmDOI{http://dx.doi.org/10.1145/3093742.3093927}
\acmISBN{978-1-4503-5065-5/17/06}
\acmConference{DEBS '17}{June 19-23, 2017}{Barcelona, Spain}
\acmYear{2017}
\copyrightyear{2017}
\acmPrice{15.00}

\begin{document}

\title{\SYS: A Reactive Middleware Framework for Secure~Data~Stream~Processing}
\date{}

\author{Aur\'elien Havet}
\affiliation{  \institution{University of Neuchâtel}
  \city{Neuchâtel}
  \country{Switzerland}
}
\email{aurelien.havet@unine.ch}

\author{Rafael Pires}
\affiliation{  \institution{University of Neuchâtel}
  \city{Neuchâtel}
  \country{Switzerland}
}
\email{rafael.pires@unine.ch}

\author{Pascal Felber}
\affiliation{  \institution{University of Neuchâtel}
  \city{Neuchâtel}
  \country{Switzerland}
}
\email{pascal.felber@unine.ch}

\author{Marcelo Pasin}
\affiliation{  \institution{University of Neuchâtel}
  \city{Neuchâtel}
  \country{Switzerland}
}
\email{marcelo.pasin@unine.ch}

\author{Romain Rouvoy}
\affiliation{  \institution{Univ. Lille / Inria / IUF}
  \city{Lille}
  \country{France}
}
\email{romain.rouvoy@univ-lille.fr}

\author{Valerio Schiavoni}
\affiliation{  \institution{University of Neuchâtel}
  \city{Neuchâtel}
  \country{Switzerland}
}
\email{valerio.schiavoni@unine.ch}

\renewcommand{\shortauthors}{A. Havet, R. Pires, P. Felber, M. Pasin, R. Rouvoy, V. Schiavoni}

\clubpenalty=10000
\widowpenalty=10000

\begin{abstract}
The growing adoption of distributed data processing frameworks in a wide diversity of application domains challenges end-to-end integration of properties like security, in particular when considering deployments in the context of large-scale clusters or multi-tenant Cloud infrastructures.

This paper therefore introduces \SYS{}, a reactive middleware framework to deploy and process secure streams at scale.
Its design combines the high-level reactive dataflow programming paradigm with Intel{\textregistered}'s low-level \emph{software guard extensions} (SGX) in order to guarantee privacy and integrity of the processed data.
The experimental results of \SYS{} are promising: while offering a fluent scripting language based on \textsc{Lua}, our middleware delivers high processing throughput, thus enabling developers to implement secure processing pipelines in just few lines of code.
\end{abstract}


\maketitle

\section{Introduction}\label{sec:introduction}

The data deluge imposed by a world of ever-connected devices, whose most emblematic example is the \emph{Internet of Things} (IoT), has fostered the emergence of novel data analytics and processing technologies to cope with the ever increasing \emph{volume}, \emph{velocity}, and \emph{variety} of information that characterize the big data era.
In particular, to support the continuous flow of information gathered by millions of IoT devices, data streams have emerged as a suitable paradigm to process flows of data at scale.
However, as some of these data streams may convey sensitive information, stream processing requires support for end-to-end security guarantees in order to prevent third parties accessing restricted data.

This paper therefore introduces \SYS{}, our initial work on a middleware framework for developing and deploying secure stream processing on untrusted distributed environments.
\SYS{} supports the implementation, deployment, and execution of stream processing tasks in distributed settings, from large-scale clusters to multi-tenant Cloud infrastructures.
More specifically, \SYS{} adopts a message-oriented~\cite{mom} middleware, which integrates with the SSL protocol~\cite{freier2011secure} for data communication and the current version of Intel{\textregistered}'s \emph{software guard extensions} (SGX)~\cite{costan_intel} to deliver end-to-end security guarantees along data stream processing stages.
\SYS{} can scale vertically and horizontally by adding or removing processing nodes at any stage of the pipeline, for example to dynamically adjust according to the current workload.
The design of the \SYS{} system is inspired by the dataflow programming paradigm~\cite{uustalu_essence_2005}: the developer combines together several independent processing components (\emph{e.g.}, mappers, reducers, sinks, shufflers, joiners) to compose specific processing pipes.
Regarding packaging and deployment, \SYS{} smoothly integrates with industrial-grade lightweight virtualization technologies like Docker~\cite{docker}.

In this paper, we propose the following contributions: (i)~we describe the design of \SYS, (ii)~we provide details of our reference implementation, in particular on how to smoothly integrate our runtime inside an SGX enclave, and (iii) we perform an extensive evaluation with micro-benchmarks, as well as with a real-world dataset.

The remainder of the paper is organized as follows.
To better understand the design of \SYS, Section~\ref{sec:background} delivers a brief introduction to today's SGX operating mechanisms.
The architecture of \SYS{} is then introduced in Section~\ref{sec:architecture}.
Our implementation choices and an example of a \SYS{} program are reported in Section~\ref{sec:implementation}.
Section~\ref{sec:eval} discusses our extensive evaluation, presenting a detailed analysis of micro-benchmark performances, as well as more comprehensive macro-benchmarks with real-world datasets.
Some related works to this topic are gathered in Section~\ref{sec:rw}.
Finally, Section~\ref{sec:conclusion} briefly describes our future work and concludes.
 
\section{SGX Lightning Tour}\label{sec:background}

The design of \SYS{} revolves around the availability of SGX features in the host machines.
It consists in a \emph{trusted execution environment} (TEE) recently introduced into Intel{\textregistered} SkyLake, similar in spirit to ARM \textsc{TrustZone}~\cite{arm2009security} but much more powerful.
Applications create secure \emph{enclaves} to protect the integrity and the confidentiality of the data and the code being executed.

\begin{figure}[!t]
  \centering
  \includegraphics[width=\linewidth]{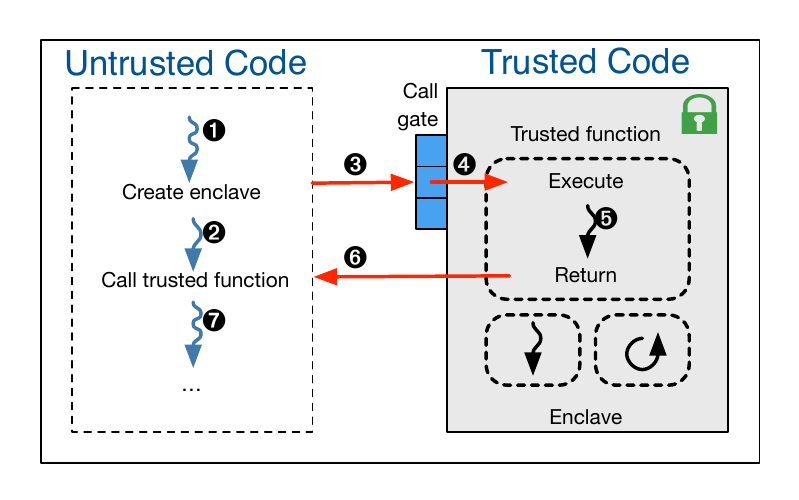}
  \caption{SGX core operating principles.}
  \label{fig:sgx}
\end{figure}

The SGX mechanism, as depicted in Figure~\ref{fig:sgx}, allows applications to access confidential data from inside the enclave.
The architecture guarantees that an attacker with physical access to a machine will not be able to tamper with the application data without being noticed.
The CPU package represents the security boundary.
Moreover, data belonging to an enclave is automatically encrypted and authenticated when stored in main memory.
A memory dump on a victim’s machine will produce encrypted data.
A \emph{remote attestation protocol} allows one to verify that an enclave runs on a genuine Intel{\textregistered} processor with SGX.
An application using enclaves must ship a signed (not encrypted) shared library (a shared object file in Linux) that can possibly be inspected by malicious attackers.

In the current version of SGX, the \emph{enclave page cache} (EPC) is a $128\,\mathit{MB}$ area of memory\footnote{Future releases of SGX might relax this limitation~\cite{mckeen2016intel}.} predefined at boot to store enclaved code and data.
At most around $90\,\mathit{MB}$ can be used by application’s memory pages, while the remaining area is used to maintain SGX metadata.
Any access to an enclave page that does not reside in the EPC triggers a page fault.
The SGX driver interacts with the CPU to choose which pages to evict.
The traffic between the CPU and the system memory is kept confidential by the \emph{memory encryption engine} (MEE)~\cite{gueron2016memory}, also in charge of tamper resistance and replay protection.
If a cache miss hits a protected region, the MEE encrypts or decrypts data before sending to, respectively fetching from, the system memory and performs integrity checks.
Data can also be persisted on stable storage protected by a seal key.
This allows the storage of certificates, waiving the need of a new remote attestation every time an enclave application restarts.

The execution flow of a program using SGX enclaves is like the following.
First, an enclave is created (see Figure~\ref{fig:sgx}-\ding{202}).
As soon as a program needs to execute a trusted function (\ding{203}), it executes SGX's primitive \texttt{ecall} (\ding{204}).
The call goes through the SGX call gate to bring the execution flow inside the enclave (\ding{205}).
Once the trusted function is executed by one of the enclave's threads (\ding{206}), its result is encrypted and sent back (\ding{207}) before giving back the control to the main processing thread (\ding{208}).
 
\section{Architecture}\label{sec:architecture}

\begin{figure*}[!t]
  \centering
  \includegraphics[width=\linewidth]{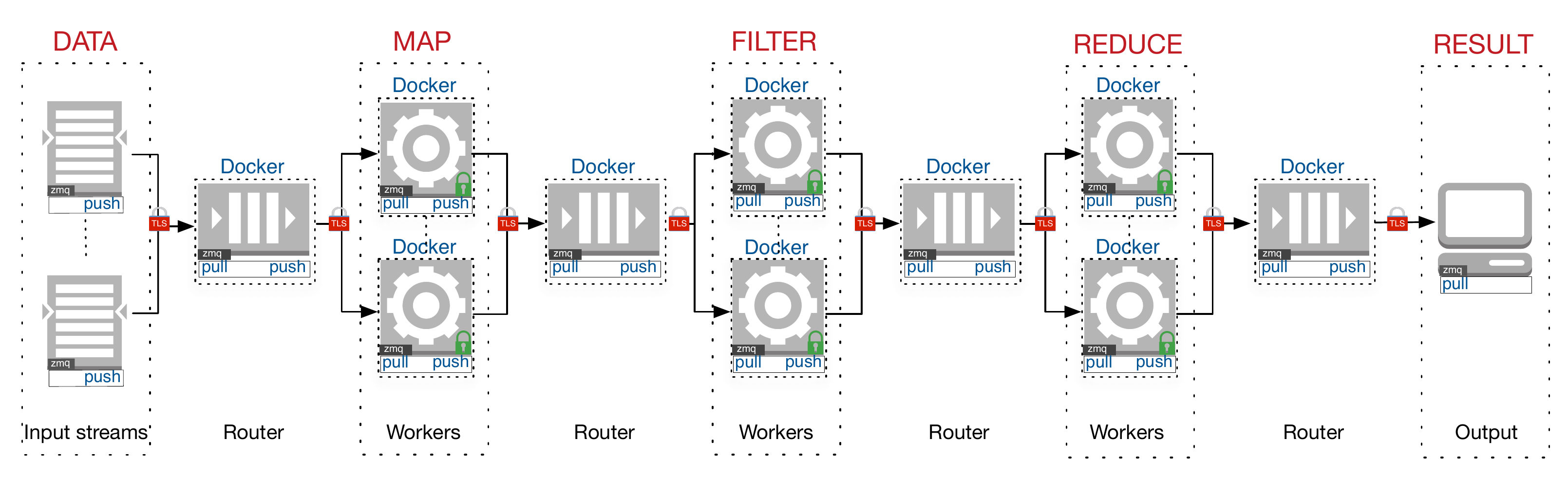}
  \caption{Example of \SYS{} pipeline architecture.}
  \label{fig:architecture_pipeline}
\end{figure*}

The architecture of \SYS{} comprises a combination of two different types of base components: \textsf{worker} and \textsf{router}.
A \textsf{worker} component continuously listens for incoming data by means of non-blocking I/O.
As soon as data flows in, an application-dependent business logic is applied.
A typical use-case is the deployment of a classic filter/map/reduce pattern from the functional programming paradigm~\cite{bird_introduction_1988}.
In such a case, worker nodes execute only one function, namely \texttt{map}, \texttt{filter}, or \texttt{reduce}.
A \textsf{router} component acts as a message broker between workers in the pipeline and transfers data between them according to a given \emph{dispatching policy}.
Figure~\ref{fig:architecture_pipeline} depicts a possible implementation of this dataflow pattern using the \SYS{} middleware.

\SYS{} is designed to support the processing of sensitive data inside SGX enclaves.
As explained in the previous section, the \emph{enclave page cache} (EPC) is currently limited to $128\,\mathit{MB}$.
To overcome this limitation, we settled on a lightweight yet efficient embeddable runtime, based on the \textsc{Lua} virtual machine (\luavm)~\cite{ierusalimschy_luaextensible_1996} and the corresponding multi-paradigm scripting language~\cite{lualang}.
The \textsc{Lua} runtime requires only few kilobytes of memory, it is designed to be embeddable, and as such it represents an ideal candidate to execute in the limited space allowed by the EPC.
Moreover, the application-specific functions can be quickly prototyped in \textsc{Lua}, and even complex algorithms can be implemented with an almost 1:1 mapping from pseudo-code~\cite{leonini2009splay}.
We provide further implementation details of the embedding of the \luavm inside an SGX enclave in Section~\ref{sec:implementation}.

Each component is wrapped inside a lightweight Linux container (in our case, the \emph{de~facto} industrial standard Docker~\cite{docker}).
Each container embeds all the required dependencies, while guaranteeing the correctness of their configuration, within an isolated and reproducible execution environment.
By doing so, a \SYS{} processing pipeline can be easily deployed without changing the source code on different public or private infrastructures.
For instance, this will allow developers to deploy \SYS{} to Amazon EC2 container service~\cite{awsec2container}, where SkyLake-enabled instances will soon be made available~\cite{amazonskylake}, or similarly to Google compute engine~\cite{gceskylake}.
The deployment of the containers can be transparently executed on a single machine or a cluster, using a Docker network and the Docker Swarm scheduler~\cite{docker:swarm_2016}.

The communication between workers and routers leverages \zmq{}, a high-performance asynchronous messaging library~\cite{zero_mq}.
Each router component hosts inbound and outbound queues.
In particular, the routers use the \zmq's pipeline pattern~\cite{zero_mq:pipeline} with the \textsc{Push}-\textsc{Pull} socket types.

The inbound queue is a \textsc{Pull} socket.
The messages are streamed from a set of anonymous\footnote{\emph{Anonymous} refers to a peer without any identity: the server socket ignores which worker sent the message.} \textsc{Push} peers (\emph{e.g.}, the upstream workers in the pipeline).
The inbound queue uses a fair-queuing scheduling to deliver the message to the upper layer.
Conversely, the outbound queue is a \textsc{Push} socket, sending messages using a round-robin algorithm to a set of anonymous \textsc{Pull} peers---\emph{e.g.}, the downstream workers.

This design allows us to dynamically scale up and down each stage of the pipeline in order to adapt it to application's needs or the workload.
Finally, \zmq{} guarantees that the messages are delivered across each stage via reliable TCP channels.

We define the processing pipeline components and their chaining by means of Docker's Compose~\cite{docker:compose} description language.
Listing~\ref{pipeline-desc} reports on a snippet of the description used to deploy the architecture in Figure~\ref{fig:architecture_pipeline}.
Once the processing pipeline is defined, the containers must be deployed on the computing infrastructure.
We exploit the \texttt{constraint} placement mechanisms to enforce the Docker Swarm's scheduler in order to deploy workers requiring SGX capabilities into appropriate hosts.
In the example, an \texttt{sgx\_mapper} nodes is deployed on an SGX host by specifying \texttt{"constraint:type==sgx"} in the Compose description.

\vspace{10pt}
\begin{lstlisting}[language=YAML,caption={\SYS pipeline examples. Some attributes (\texttt{volume}, \texttt{networks}, \texttt{env\_file}) are omitted.},label=pipeline-desc][!t]
sgx_mapper:
  image: "${IMAGE_SGX}"
  entrypoint: ./start.sh sgx-mapper.lua
  environment:
    - TO=tcp://router_mapper_filter:5557
    - FROM=tcp://router_data_mapper:5556
    - "constraint:type==sgx"
  devices:
    - "/dev/isgx"

router_data_mapper:
  image: "${IMAGE}"
  hostname: router_data_mapper
  entrypoint: lua router.lua
  environment:
    - TO=tcp://*:5556
    - FROM=tcp://*:5555
    - "constraint:type==sgx"

data_stream:
  image: "${IMAGE}"
  entrypoint: lua data-stream.lua
  environment:
    - TO=tcp://router_data_mapper:5555
    - "constraint:type==sgx"
    - DATA_FILE=the_stream.csv
\end{lstlisting}
 
\section{Implementation Details}\label{sec:implementation}

\SYS{} is implemented in \textsc{Lua} (v5.3).
The implementation of the middleware itself requires careful engineering, especially with respect to the integration in the SGX enclaves (explained later).
However, a \SYS{} use-case can be implemented in remarkably few lines of code.
For instance, the implementation of the map/filter/reduce accounts for only $120$ lines of code (without counting the dependencies).
The framework partially extends \rxl~\cite{github:rxlua}, a library for reactive programming in \textsc{Lua}.
\rxl provides to the developer the required API to design a data stream processing pipeline following a dataflow programming pattern~\cite{uustalu_essence_2005}.

Listing~\ref{pipeline-example} provides an example of a \rxl program (and consequently a \SYS{} program) to compute the average age of a population by chaining \texttt{:map}, \texttt{:filter}, and \texttt{:reduce} functions.\footnote{Note that in our evaluation the code executed by each worker is confined into its own \textsc{Lua} file.}
The \texttt{:subscribe} function performs the subscription of 3 functions to the data stream.
Following the \emph{observer} design pattern~\cite{szallies_using_1997}, these functions are observers, while the data stream is an observable.

\begin{lstlisting}[language=LUA,caption={Example of process pipeline with RxLua.},label=pipeline-example]
Rx.Observable.fromTable(people)
 :map(
   function(person)
     return person.age
   end
 )
 :filter(
   function(age)
     return age > 18
   end
 )
 :reduce(
   function(accumulator, age)
     accumulator[count] = (accumulator.count
       or 0) + 1
     accumulator[sum] = (accumulator.sum
       or 0) + age
     return accumulator
   end
 )
 :subscribe(
   function(datas)
     print("Adult people average:",
       datas.sum / datas.count)
   end,
   function(err)
     print(err)
   end,
   function()
     print("Process complete!")
   end
 )
\end{lstlisting}

\SYS{} dynamically ships the business logic for each component into a dedicated Docker container and executes it.
The communication between the Docker containers (the router and the worker components) happens through \zmq (v4.1.2) and the corresponding \textsc{Lua} bindings~\cite{github:lzmq}.
Basically, \SYS{} abstracts the underlying network and computing infrastructure from the developer, by relying on \zmq and Docker.

Under the SGX threat model where the system software is completely untrusted, system calls are not allowed inside secure enclaves.
As a consequence, porting a legacy application or runtime, such as the \textsc{Lua} interpreter, is challenging.
To achieve this task, we traced all system calls made by the interpreter to the standard C library and replaced them by alternative implementations that either mimic the real behavior or discard the call.
Our changes to the vanilla \textsc{Lua} source code consist of the addition of about $600$ lines of code, or $2.5\,\mathit{\%}$ of its total size.
By doing so, \textsc{Lua} programs operating on files, network sockets or any other input/output device do not execute as they normally do outside the enclaves. This inherent SGX limitation also reinforces the system security guarantees offered to the application developers.
The \SYS{} framework safely ships the data and code to enclaves.
Hence, the \textsc{Lua} scripts executed within the SGX enclave do not use (read/write) files or sockets.
Wrapper functions are nevertheless installed in the SGX-enabled \luavm to prevent any of such attempts.

An additional constraint imposed by the secure SGX enclaves is the impossibility of dynamically linking code.
The reason is that the assurance that a given code is running inside a SGX-enabled processor is made through the measurement of its content when the enclave is created.
More specifically, this measurement is the result of \texttt{EREPORT} instruction, an SGX-specific report that computes a cryptographically secure hash of code, data and a few data structures, which overall builds a snapshot of the state of the enclave (including threads, memory heap size, etc.) and the processor (security version numbers, keys, etc.).
Allowing more code to be linked dynamically at runtime would break the assurance given by the attestation mechanism on the integrity of the code being executed, allowing for example an attacker to load a malicious library inside the enclave.

In the case of \textsc{Lua}, a direct consequence is the impossibility of loading \textsc{Lua} extensions using the traditional dynamic linking technique.
Every extension has to be statically compiled and packed with the enclave code.
To ease the development of \SYS applications, we statically compiled \emph{json}~\cite{rfc7159}, and \emph{csv}~\cite{rfc4180} parsers within our enclaved \textsc{Lua} interpreter.
With these libraries, the size of the VM and the complete runtime still remains reasonably small, approximately $220\,\mathit{KB}$ ($19\,\mathit{\%}$ larger than the original).

\begin{figure}[t!]
  \centering
  \includegraphics[width=.9\linewidth]{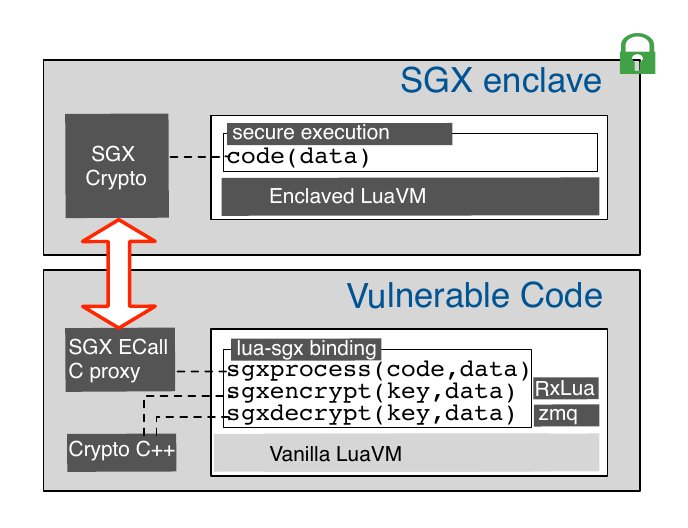}
  \caption{Integration between \textsc{Lua} and Intel{\textregistered} SGX.}
  \label{fig:arch-luasgx}
\end{figure}

While this restricted \textsc{Lua} has been adapted to run inside SGX enclaves, we still had to provide a support for communications and the reactive streams framework itself.
To do so, we use an external vanilla \textsc{Lua} interpreter, with a couple adaptations that allowed the interaction with the SGX enclaves and the \luavm therein.
Figure~\ref{fig:arch-luasgx} shows the resulting architecture.
We extend the \textsc{Lua} interface with 3 functions: \texttt{sgxprocess}, \texttt{sgxencrypt}, and \texttt{sgxdecrypt}.
The first one forwards the encrypted code and data to be processed in the enclave, while the remaining two provide cryptographic functionalities.
In this work, we assume that attestation and key establishment was previously performed.
As a result, keys safely reside within the enclave.
We plan to release our implementation as open-source.\footnote{https://github.com/vschiavoni/SecureStreams-DEBS17}
 
\section{Evaluation}\label{sec:eval}

This section reports on our extensive evaluation of \SYS{}.
First, we present our evaluation settings.
Then, we describe the real-world dataset used in our macro-benchmark experiments.
We then dig into a set of micro-benchmarks that evaluate the overhead of running the \luavm inside the SGX enclaves.
Finally, we deploy a full \SYS{} pipeline, scaling the number of workers per stage, to study the limits of the system in terms of throughput and scalability.

\subsection{Evaluation Settings}

We have experimented on machines using a Intel{\textregistered} Core\texttrademark~i7-6700 processor~\cite{intel:i7_6700} and $8\,\mathit{GiB}$ RAM.
We use a cluster of 2 machines based on \textsc{Ubuntu} 14.04.1 LTS (kernel 4.2.0-42-generic).
The choice of the Linux distribution is driven by compatibility reasons with the Intel{\textregistered} SGX SDK (v1.6).
The machines run Docker (v1.13.0) and each node joins a Docker Swarm~\cite{docker:swarm_2016} (v1.2.5) using the Consul~\cite{consul} (v0.5.2) discovery service.
The Swarm manager and the discovery service are deployed on a distinct machine.
Containers building the pipeline leverage the Docker overlay network to communicate to each other, while machines are physically interconnected using a switched $1\,\mathit{Gbps}$ network.

\subsection{Input Dataset}

In our experiments, we process a real-world dataset released by the \emph{American Bureau of Transportation Statistics}~\cite{rita:bts}.
The dataset reports on the flight departures and arrivals of $20$ air carriers~\cite{statistical_computing:data}.
We implement a benchmark application atop of \SYS{} to compute average delays and the total of delayed flights for each air carrier (cf. Table~\ref{tab:appsize}).
We design and implement the full processing pipeline, that (i)~parses the input datasets (in a comma-separated-value format) to data structure (\textsf{map}), (ii)~filters data by relevancy (\emph{i.e.}, if the data concerns a delayed flight), and (iii)~finally reduces it to compute the desired information.\footnote{This experiment is inspired by Kevin Webber's blog entry \emph{diving into Akka streams}: \url{https://blog.redelastic.com/diving-into-akka-streams-2770b3aeabb0}.}
We use the $4$ last years of the available dataset (from 2005 to 2008), for a total of $28$ millions of entries to process and $2.73\,\mathit{GB}$ of data.

\begin{table}[t!]
    \centering
    \begin{tabular}{l|r}
   \textbf{System layer}          & \textbf{Size (LoC)} \\
\hline
\textsc{DelayedFlights} app    & $86$ \\
\textsc{SecureStreams} library & $350$ \\
\textsc{RxLua} runtime         & $1,481$ \\
\hline
\hline
Total                          & $1,917$ \\
      \end{tabular}
    \caption{Benchmark app based on \SYS{}.}
  \label{tab:appsize}
\end{table}

\subsection{Micro-Benchmark: \textsc{Lua} in SGX}

We begin our evaluation with a set of micro-benchmarks to evaluate performance of the integration of the \luavm inside the SGX enclaves.
First, we estimate the cost of execution for functions inside the enclave.
This test averages the execution time of 1 million function calls, without any data transfer.
We compare against the same result without SGX.
While non-enclaved function calls took $23.6\,\mathit{ns}$, the performances inside the enclave drop down to on average $2.35\,\mathit{s}$---\textit{i.e.}, approximately two orders of magnitude worse.
We then assess the cost of copying data from the unshielded execution to the enclave and we compare it with the time required to compute the same on the native system.
We initialize a buffer of $100\,\mathit{MB}$ with random data and copy its content inside the enclave.
The data is split into chunks of increasing sizes.
Our test executes one function call to transfer each chunk, until all data is transfered.
Each point in the plot corresponds to the average of $20$ runs.
Correctness of the copies was verified by \textsf{SHA256} digest comparison between reproduced memory areas.

\begin{figure}[t!]
  \centering
  \includegraphics[width=\linewidth]{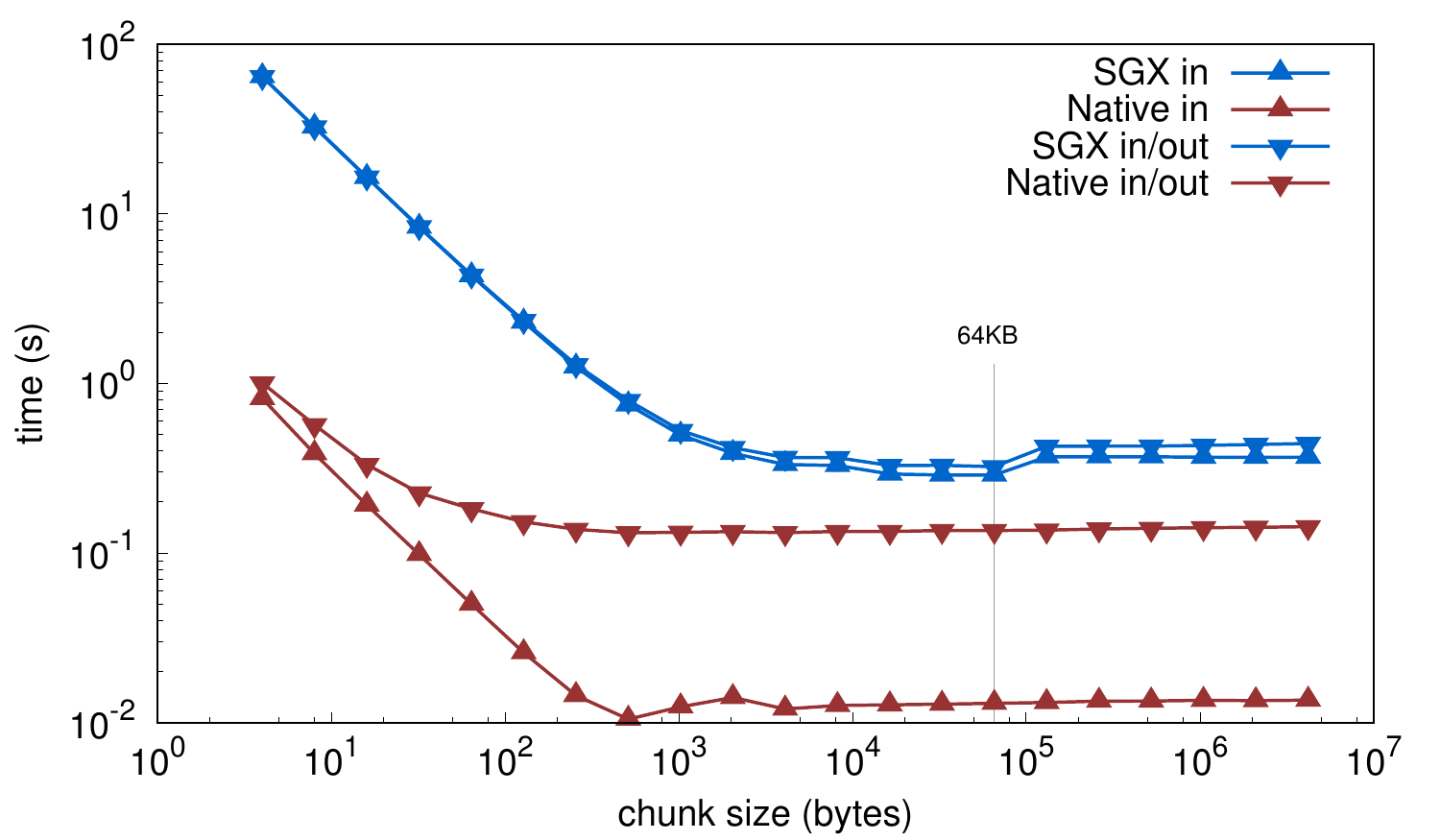}
  \caption{Execution time to copy $100\,\mathit{MB}$ of memory inside an SGX enclave (\emph{in}) or to copy it back outside {\emph{in/out}.} }
  \label{fig:sgxmemcpy}
\end{figure}

Figure~\ref{fig:sgxmemcpy} shows the results for $4$ different variants, comparing the native and the SGX version to only copy the data inside the enclave (\emph{in}) or to copy it inside and copying it back (\emph{in/out}).
When using smaller chunks, the function call overhead plays an important role in the total execution time.
Moreover, we notice that the call overhead steadily drops until the chunk size reaches the size of $64\,\mathit{KB}$ (vertical line).
We can also notice that copying data back to non-SGX execution imposes an overhead of at most $20\,\mathit{\%}$ when compared to the one-way copy.
These initial results are used as guidelines to drive the configuration of the streaming pipeline, in particular with respect to the size of the chunks exchanged between the processing stages. The larger the chunks, the smaller the overhead induced by the transfer of data within the SGX enclave.

\begin{figure}[t!]
  \centering
  \includegraphics[width=\linewidth]{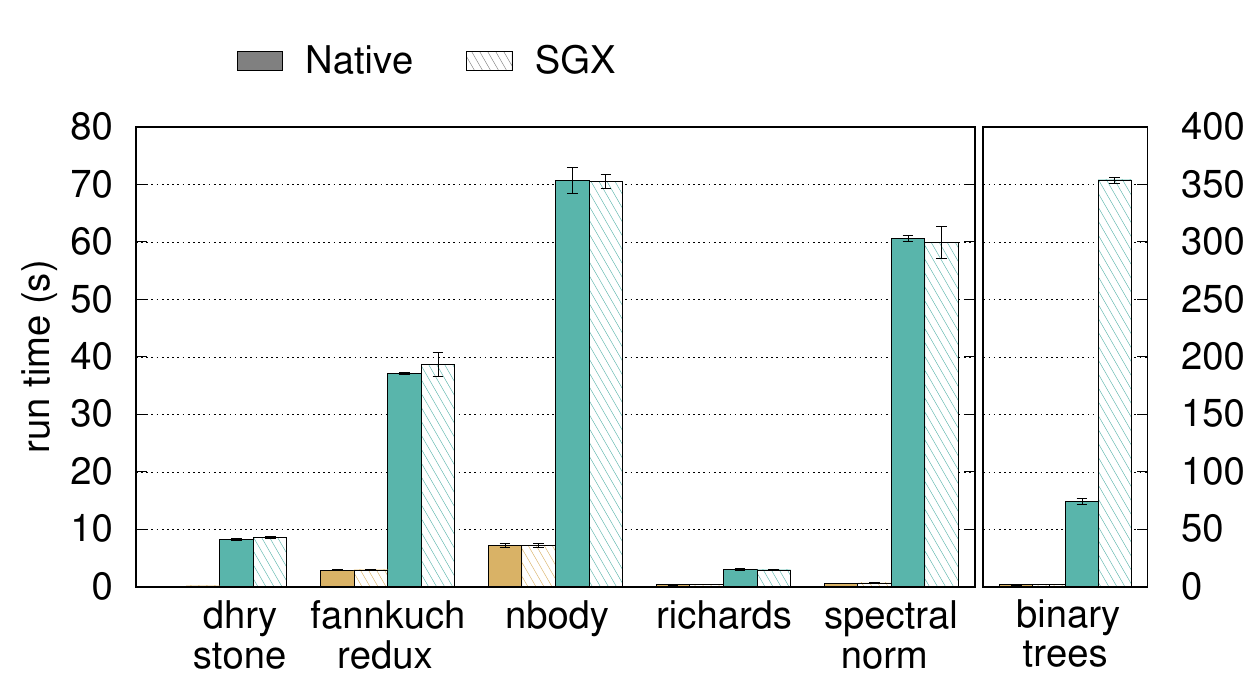}
  \caption{Enclave versus native running times for \textsc{Lua} benchmarks.}
  \label{fig:luabenchs}
\end{figure}

Once the data and the code are copied inside the enclave, the \luavm must indeed execute the code before returning the control.
Hence, we evaluate here the raw performances of the enclaved SGX \luavm.
We select $6$ available benchmarks from a standard suite of tests~\cite{bolz2015}.
We based this choice on their library dependencies (by selecting the most standalone ones) and the number of input/output instructions they execute (selecting those with the fewest I/O).
Each benchmark runs $20$ times with the same pair of parameters of the original paper, shown in the even and odd lines of Table~\ref{tab:luabmarks}.
Figure~\ref{fig:luabenchs} depicts the total time (average and standard deviation) required to complete the execution of the $6$ benchmarks.
We use a bar chart plot, where we compare the results of the \emph{Native} and \emph{SGX} modes.
For each of the $6$ benchmarks, we present two bars next to each other (one per executing mode) to indicate the different configuration parameters used.
Finally, for the sake of readability, we use a different y-axis scale for the \textsf{binarytrees} case (from $0$ to $400$\,s), on the right-side of the figure.

\newcommand{\higparamcolor}{\rowcolor[rgb]{0.79,0.91,0.90}\cellcolor{white}}
\newcommand{\lowparamcolor}{\rowcolor[rgb]{0.94,0.88,0.76}\cellcolor{white}}
\begin{table}[t!]
    \centering
    \begin{tabular}{r|c|c|c}
                          &configuration &memory      &ratio \\
                       &parameter     &peak        &SGX/Native \\
\hline
\lowparamcolor
\textsf{dhrystone}     &50\,K      &275\,KB       & 1.14 \\
\higparamcolor
                       &5\,M       &275\,KB       & 1.04 \\
\hline
\lowparamcolor
\textsf{fannkuchredux} &10         &28\,KB        & 0.99 \\
\higparamcolor
                       &11         &28\,KB        & 1.04 \\
\hline
\lowparamcolor
\textsf{nbody}         &2.5\,M     &38\,KB        & 0.99 \\
\higparamcolor
                       &25\,M      &38\,KB        & 1.00 \\
\hline
\lowparamcolor
\textsf{richards}      &10         &106\,KB       & 1.02 \\
\higparamcolor
                       &100        &191\,KB       & 0.97 \\
\hline
\lowparamcolor
\textsf{spectralnorm}  &500        &52\,KB        & 1.00 \\
\higparamcolor
                       &5\,K       &404\,KB       & 0.99 \\
\hline
\lowparamcolor
\textsf{binarytrees}   &14         &25\,MB        & 1.18 \\
\higparamcolor
                       &19         &664\,MB       & 4.76 \\
      \end{tabular}
    \caption{Parameters and memory usage for \textsc{Lua} benchmarks.}
  \label{tab:luabmarks}
\end{table}

We note that, in the current version of SGX, it is required to pre-allocate all the memory area to be used by the enclave.
The most memory-eager test (\textsf{binarytrees}) used more than $600\,\mathit{MB}$ of memory, hence using the wall clock time comparison would not be fair for smaller tests.
In such cases, almost the whole execution time is dedicated to memory allocation.
Because of that, we subtracted the allocation time from the measurements of enclave executions, based on the average for the $20$ runs.
Fluctuations on this measurement produced slight variations in the execution times, sometimes producing the unexpected result of having SGX executions faster than native ones (by at most $3\,\%$).
Table~\ref{tab:luabmarks} lists the parameters along with the maximum amount of memory used and the ratio between runtimes of SGX and Native executions.
When the memory usage is low, the ratio between the Native and SGX versions is small---\emph{e.g.}, less than $15$\,\% in our experiments.
However, when the amount of memory usage increases, performance drops to almost $5\times$ worse, as reflected in the case of the \emph{binarytrees} experiment.
The smaller the memory usage, the better performance we can obtain from SGX enclaves.

\vspace{10pt}\noindent\textbf{Synthesis.}
To conclude this series of micro-benchmarks, taming the overhead of secured executions based on SGX requires balancing the size of the chunks transfered to the enclave with the memory usage within this enclave.
In the context of stream processing systems, \SYS{} therefore uses reactive programming principles to balance the load within processing stages in order to minimize the execution overhead.

\subsection{Benchmark: Streaming Throughput}

The previous set of experiments allowed us to verify that our design, implementation, and the integration of the \luavm into the SGX enclaves is sound.
Next, we deploy a \SYS{} pipeline which includes mappers, filters and reducers.
To measure the achievable throughput of our system, as well the network overhead of our architecture, we deploy the \SYS{} pipeline in 3 different configurations.
In each case, the setup of the pipeline architecture, \emph{i.e.} the creation of the set of containers, has been done in $11\,\mathit{s}$ for the lightest configuration, in $15\,\mathit{s}$ for the heaviest one.

The first configuration allows the streaming framework to blindly bypass the SGX enclaves.
Further, it does not encrypt the input dataset before injecting it into the pipeline.
This mode operates as the baseline, yet completely \emph{unsafe}, processing pipeline.
The second mode encrypts the dataset but lets the encrypted packets skip the SGX enclaves.
This configuration requires the deployers to trust the infrastructure operator.
Finally, we deploy a fully secure pipeline, where the input dataset is encrypted and the data processing is operated inside the enclaves.
The data nodes inject the dataset, split into 4 equally-sized parts, as fast as possible.
We gather bandwidth measurements by exploiting Docker's internal monitoring and statistical module.

\begin{figure*}[!t]
\centering
\begin{tabular}{cccc}
\subfloat[Streaming throughput. Data in clear text, no SGX, 1 worker per stage.]{
  \includegraphics[width=.31\linewidth]{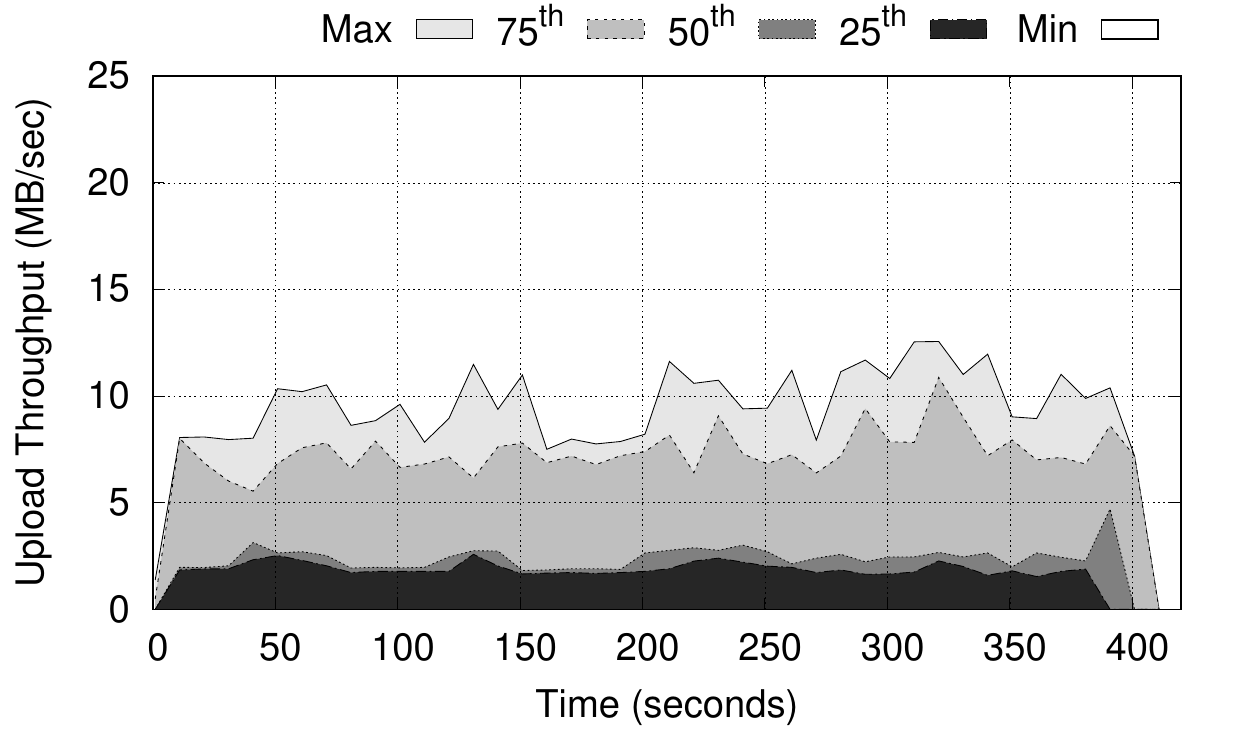}
} &
\subfloat[Streaming throughput. Encrypted data, no SGX, 1 worker per stage.]{
  \includegraphics[width=.31\linewidth]{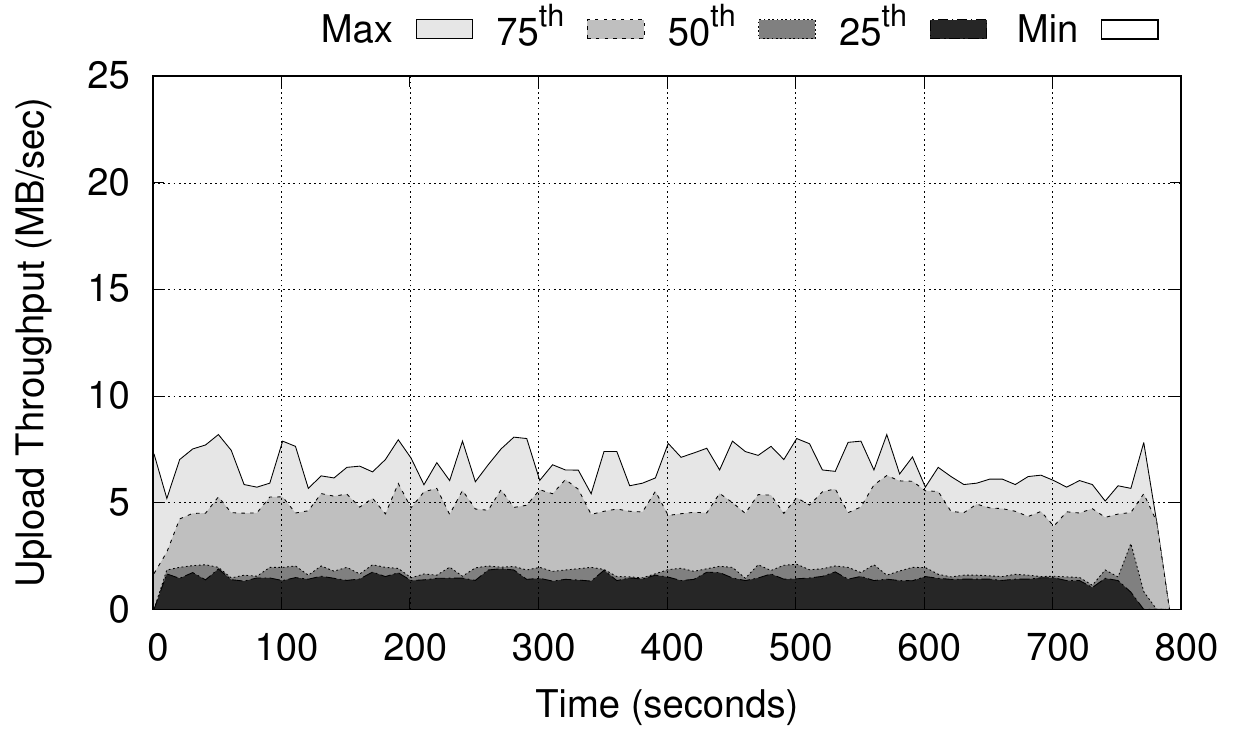}
} &
\subfloat[Streaming throughput. Encrypted data, processing SGX, 1 worker per stage.]{
  \includegraphics[width=.31\linewidth]{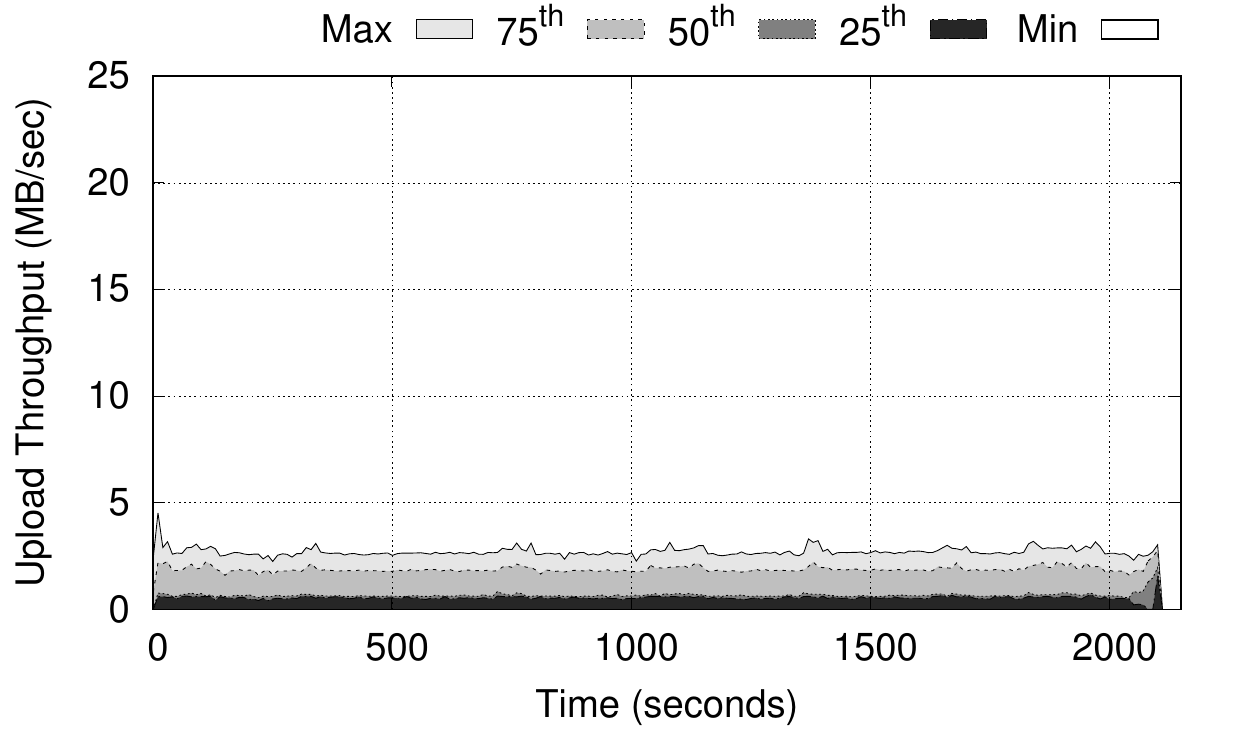}
}\cr
\subfloat[Streaming throughput. Data in clear text, no SGX, 2 workers per stage.]{
  \includegraphics[width=.31\linewidth]{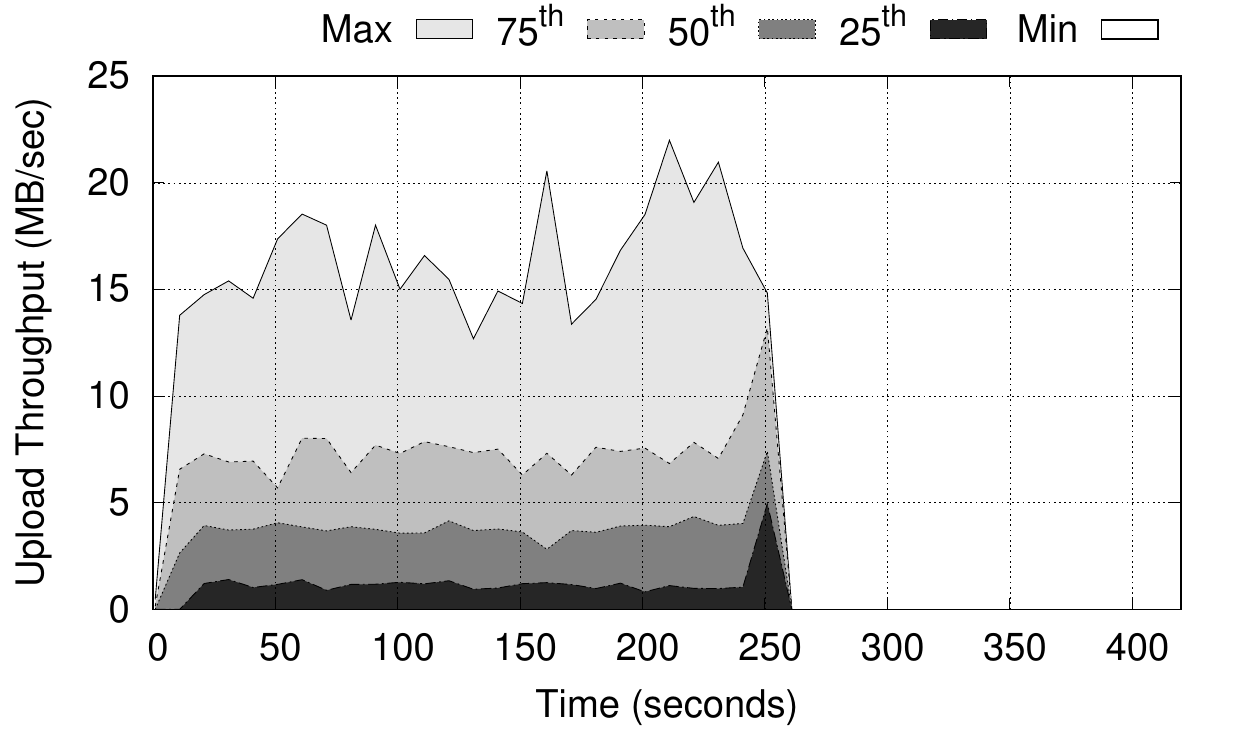}
} &
\subfloat[Streaming throughput. Encrypted data, no SGX, 2 workers per stage.]{
  \includegraphics[width=.31\linewidth]{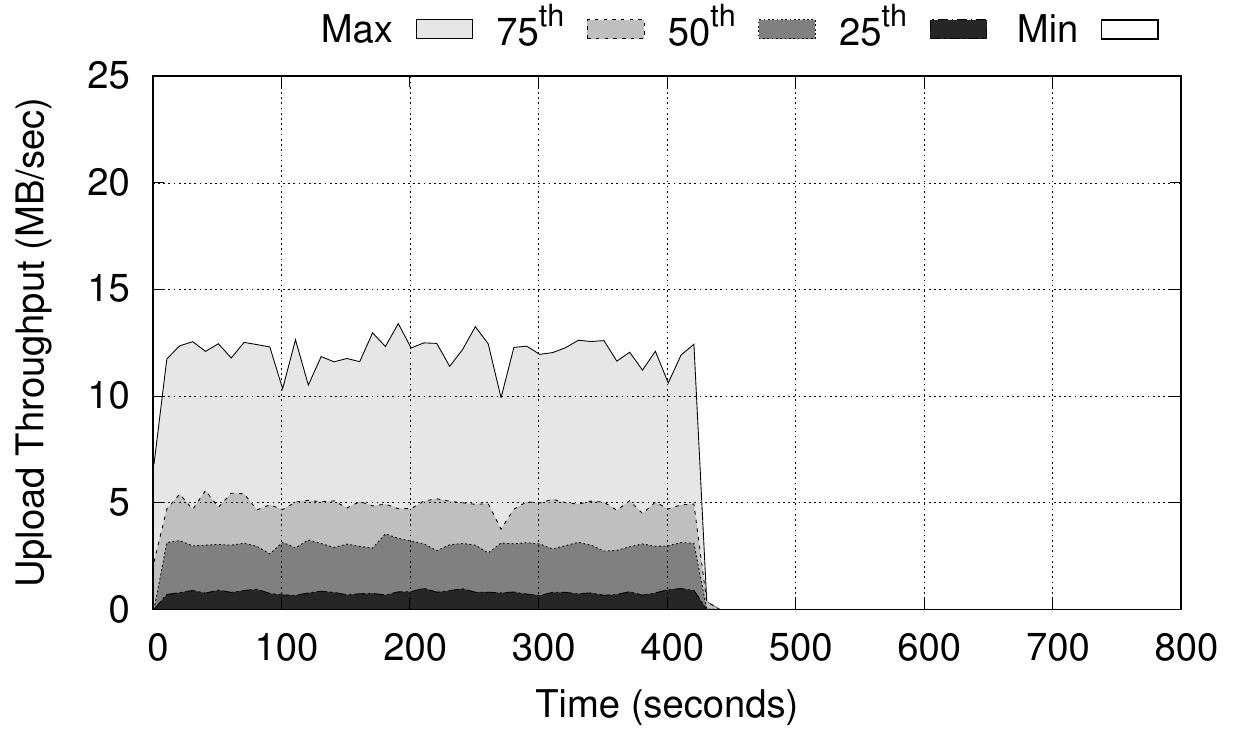}
} &
\subfloat[Streaming throughput. Encrypted data, processing SGX, 2 workers per stage.]{
  \includegraphics[width=.31\linewidth]{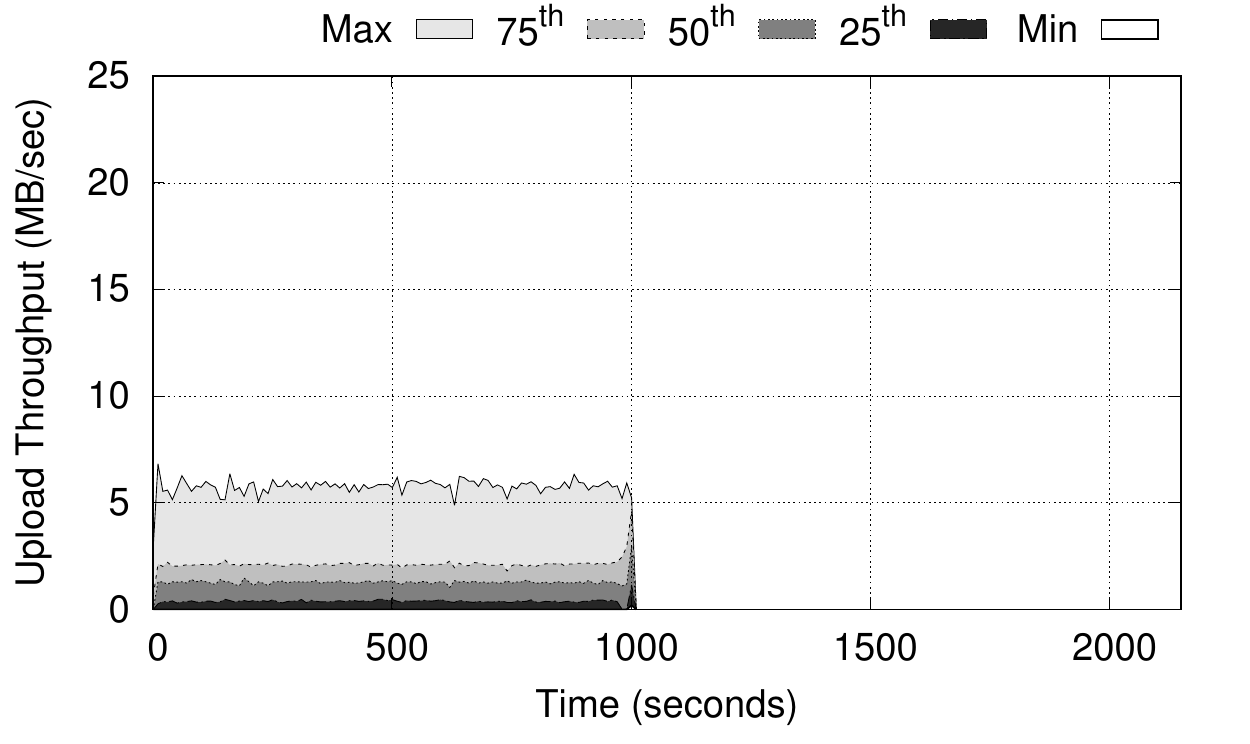}
}\cr
\subfloat[Streaming throughput. Data in clear text, no SGX, 4 workers per stage.]{
  \includegraphics[width=.31\linewidth]{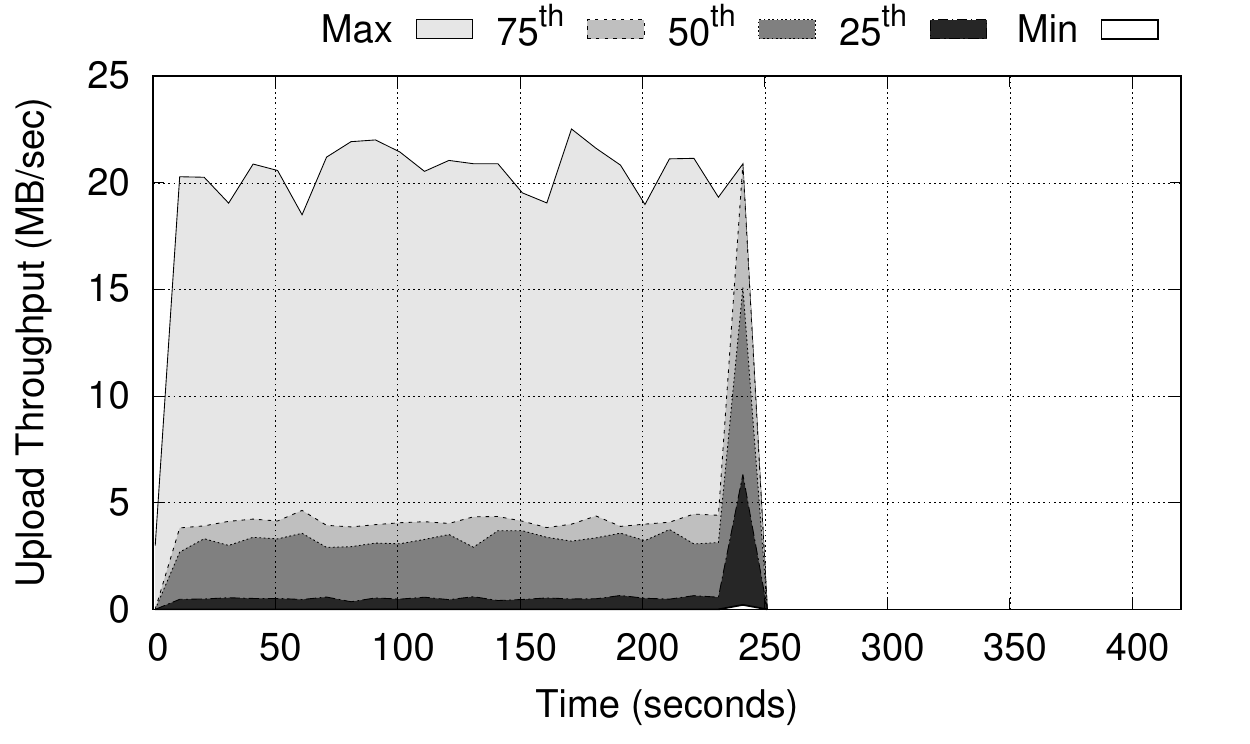}
} &
\subfloat[Streaming throughput. Encrypted data, no SGX, 4 workers per stage.]{
  \includegraphics[width=.31\linewidth]{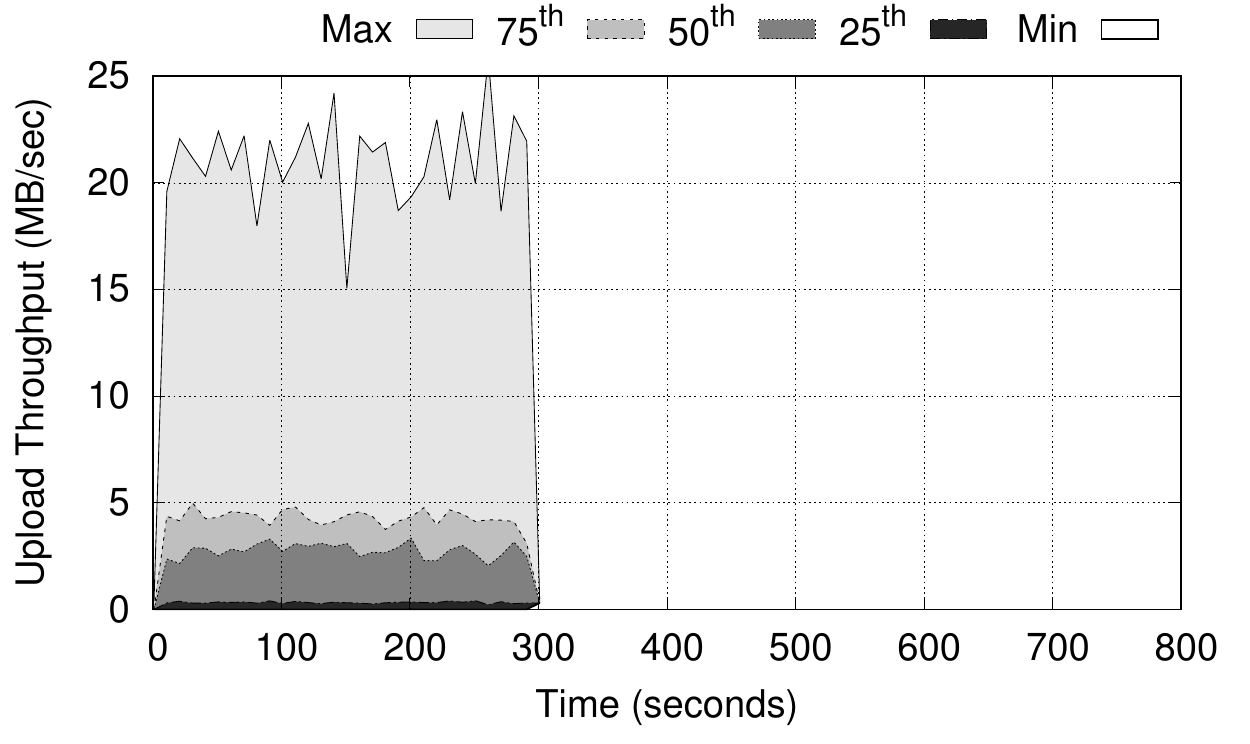}
} &
\subfloat[Streaming throughput. Encrypted data, processing SGX, 4 workers per stage.]{
  \includegraphics[width=.31\linewidth]{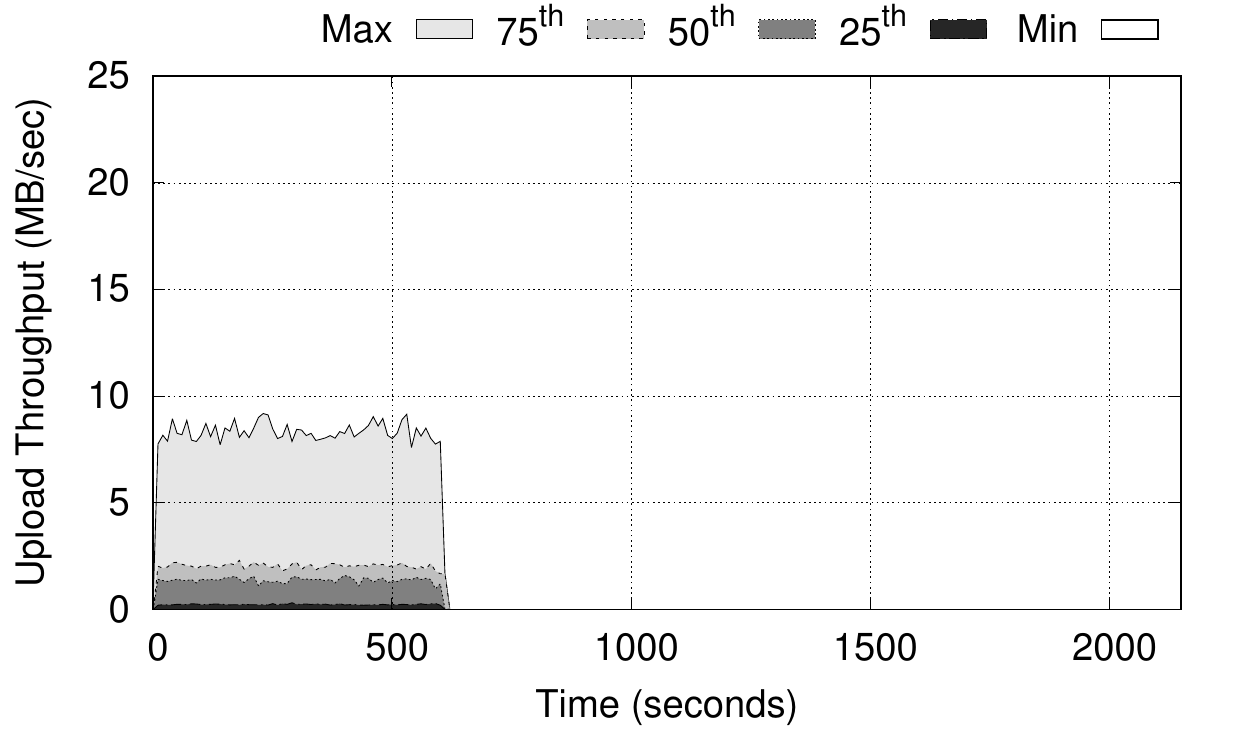}
}
\end{tabular}
\caption{Throughput comparison between normal processing (with cleartext data and no encryption), encrypted data but without enclaves, and with encrypted data and SGX processing. We scale the number of worker nodes per stage from $1$ (left-most column), $2$ (center colum) and $4$ (right-most column).}
\label{fig:throughput}
\end{figure*}

The results of these deployments are presented in Figure~\ref{fig:throughput}.
For each of the mentioned configurations, we also vary the number of workers per stage, from one (Figure~\ref{fig:throughput}-a,d,g), two (Figure~\ref{fig:throughput}-b,e,h), or four (the remaining ones.)
We use a representation based on stacked percentiles.
The white bar at the bottom represents the minimum value, the pale grey on top the maximal value.
Intermediate shades of grey represent the 25th-, 50th–, and 75th-percentiles.
For instance, in Figure~\ref{fig:throughput}-a (our baseline) the median throughput at $200\,\mathit{s}$ into the experiment almost hits $7.5\,\mathit{MB/s}$, meaning that $50\,\mathit{\%}$ of the nodes in that moment are outputting data at $7.5\,\mathit{MB/s}$ or less.
The baseline configuration, with only $1$ worker per stage, completes in $420\,\mathit{s}$, with a peak of $12\,\mathit{MB/s}$.
By doubling the number of workers reduces the processing time down to $250\,\mathit{s}$ (Figure~\ref{fig:throughput}-d), a speed-up of $41\,\mathit{\%}$.
Scaling up the workers to 4 in the baseline configuration (Figure~\ref{fig:throughput}-g) did not produce a similar speed-up.

As we start injecting encrypted datasets (Figure~\ref{fig:throughput}-b and follow-up configurations with 2 and 4 workers), the processing time almost doubles ($795\,\mathit{s}$).
The processing of the dataset is done after the messages are decrypted.
We also pay a penalty in terms of overall throughput---\emph{i.e.}, the median value rarely exceeds $5\,\mathit{MB/s}$.
On the other hand, now we observe substantial speed-ups when increasing the workers per stage, down to $430\,\mathit{s}$ and $300\,\mathit{s}$ with $2$ and $4$ workers, respectively.

The deployment of the most secure set of configurations (right-most column of plots in Figure~\ref{fig:throughput}) shows that when using encrypted datasets and executing the stream processing inside SGX enclaves one must expect longer processing times and lower throughputs.
This is the (expected) price to pay for higher-security guarantees across the full processing pipeline.
Nevertheless, one can observe that the more workers the less penalty is imposed by the end-to-end security guarantees provided by \SYS{}.

\begin{figure}[t!]
  \centering  \includegraphics[width=\linewidth]{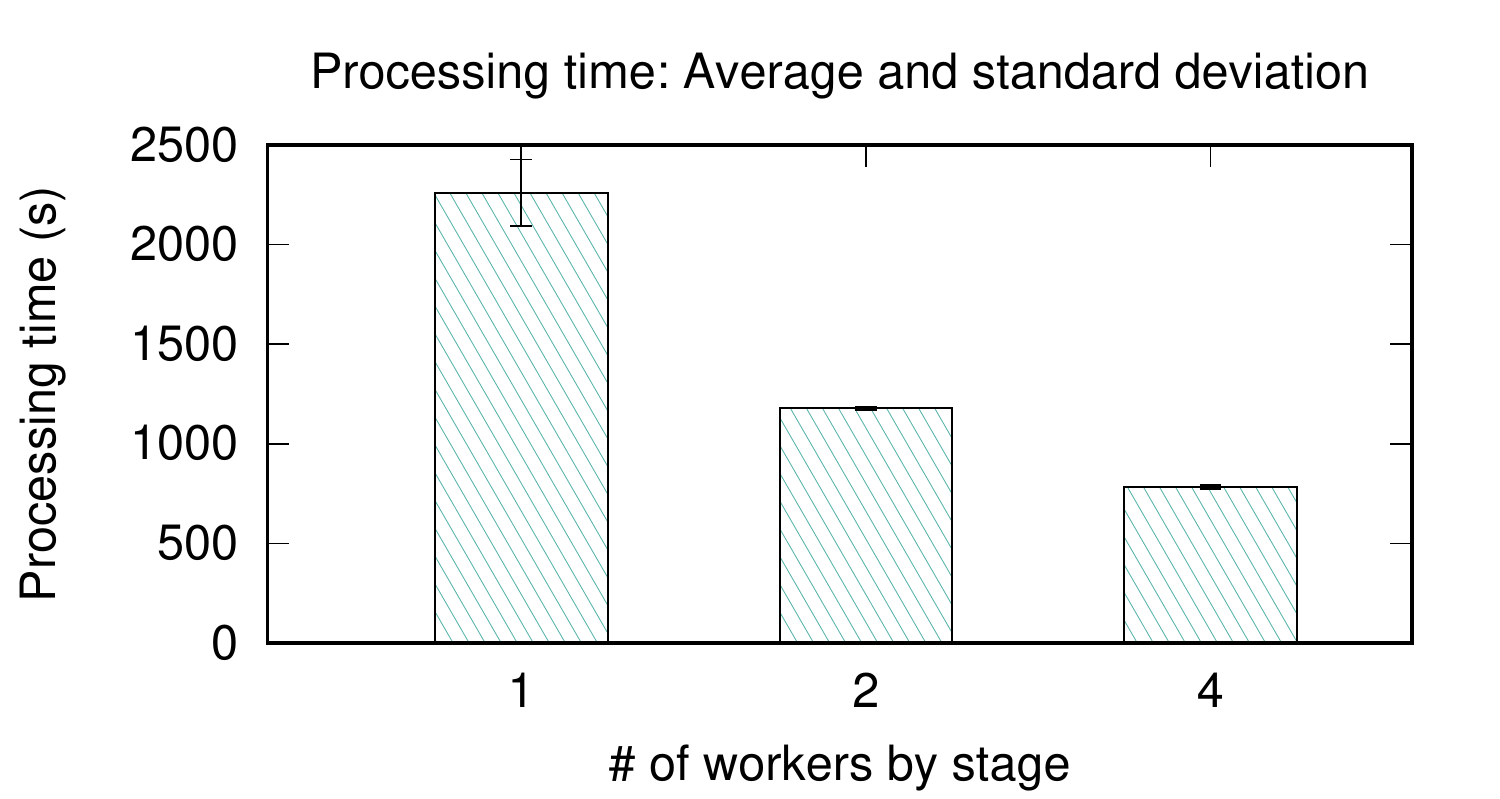}
  \caption{Scalability: processing time, average and standard deviation. The experiment is repeated 5 times, with a variation on the number of workers for each stage, each worker using SGX.}
  \label{fig:scalability:sgxfull}
\end{figure}
\begin{figure}[t!]
  \centering  \includegraphics[width=\linewidth]{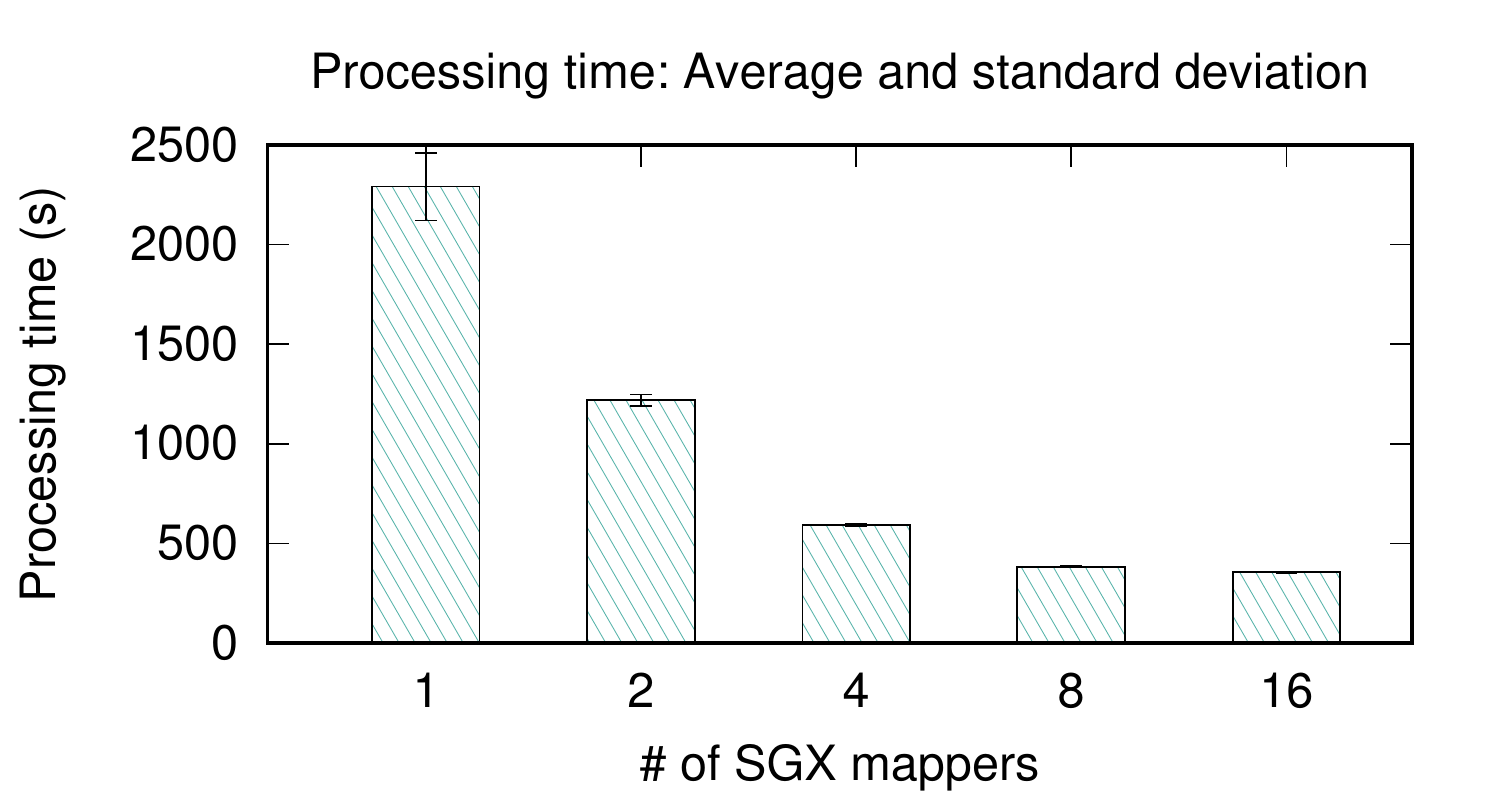}
  \caption{Scalability: processing time, average and standard deviation. The experiment is repeated 5 times, with a variation on the number of mappers SGX, other workers---1 filter worker and 1 reduce worker---do not use SGX.}
  \label{fig:scalability:sgxmapper}
\end{figure}

\subsection{Benchmark: Workers' Scalability}
To conclude our evaluation, we study \SYS{} in terms of scalability.
We consider a pipeline scenario similar to Figure~\ref{fig:architecture_pipeline} with some variations in the number of workers deployed for each stage.
We do so to better understand to what extents the underlying container scheduling system can exploit the hardware resources at its disposal.

First, we increase the number of workers for each stage of the pipeline, from $1$ to $4$.
For each of the configurations, the experiment is repeated $5$ times.
We present average and standard deviation of the total completion time to process the full dataset in Figure~\ref{fig:scalability:sgxfull}.
As expected, we observe ideal speed-up from a configuration using $1$ worker to that using $2$ workers.
However, in the configuration using $4$ workers by stage, we do not reach the same acceleration.
We explain this because, in this latter case, the number of deployed containers (which equals the sum of input data streams, workers, and routers, hence $20$ containers) is greater than the number of physical cores of the hosts ($8$ for each of the $2$ hosts used in our deployment---\emph{i.e.}, $16$ cores on our evaluation cluster).

We also study the total completion time while increasing only the number of mapper workers in the first stage of the pipeline (which we identified as the one consuming most resources) from $1$ to $16$ and maintaining the numbers of filters and reducers in the following stages constant.
As in the previous benchmark, the experiment is repeated $5$ times for each configuration and we measure the average and standard deviation of the total completion time.
Figure~\ref{fig:scalability:sgxmapper} presents the results.
Here again, we observe ideal speed-up until the number of deployed containers reaches the number of physical cores.
Beyond this number, we do not observe further improvements.
These two experiments clearly show that the scalability of \SYS{} according the number of deployed workers across the cluster is primarily limited by the total number of physical cores available.

Apart from this scalability limitation, there are other factors that reduce the observed streaming throughput, with or without involving the SGX enclaves.
For instance, our throughout experiments highlight that the system does not manage to saturate the available network bandwidth in all cases.
We believe this behaviour can be explained by the lack of optimizations in the application logic as well as possible tuning options of the inner \zmq{} queues.

As part of our future work, we therefore plan to further investigate these effects and to build on this knowledge to only scale the appropriate workers in order to maximize the overall speed-up of the deployed application.
In particular, we intend to leverage the elasticity of workers at runtime in order to cope with the memory constraints imposed by SGX and the configuration of the underlying hardware architecture, for each of the available nodes, in order to offer the best performances for secured execution of data stream processing applications built atop of \SYS{}.
 
\section{Related Work}\label{sec:rw}

Spark~\cite{Zaharia:2013:DSF:2517349.2522737} has recently gained a lot of traction as prominent solution to implement efficient stream processing.
It leverages Resilient Distributed Datasets (RDD) to provide a uniform view on the data to process.
Despite its popularity, Spark only handles unencrypted data and hence does not offer security guarantees.
Recent proposals~\cite{7840754} study possible software solutions to overcome this limitation.

Several big industrial players introduced their own stream processing solutions.
These systems are mainly used to ingest massive amounts of data and efficiently perform (real-time) analytics.
Twitter's Heron~\cite{Kulkarni:2015:THS:2723372.2742788}, and Google's Cloud DataFlow~\cite{Akidau:2015:DMP:2824032.2824076} are two prominent examples.
These systems are typically deployed on the provider's premises and are not offered \emph{as a service} to end-users.

A few dedicated solutions exist today for distributed stream processing using reactive programming.
For instance, \textsc{Reactive Kafka}~\cite{reactivekafka} allows stream processing atop of Apache \textsc{Kafka}~\cite{apachekafka,kreps2011kafka}.
These solutions do not, however, support secure execution in a trusted execution environment.

More recently, some open-source middleware frameworks (\emph{e.g.}, Apache \textsc{Spark}~\cite{apachesparkstreaming}, Apache \textsc{Storm}~\cite{apachestorm}, \textsc{Infinispan}~\cite{infinispan}) introduced APIs to allow developers to quickly set up and deploy stream processing infrastructures.
These systems rely on the \emph{Java} virtual machine (JVM)~\cite{lindholm2014java}.
However, SGX currently imposes a hard memory limit of 128\,MB to the enclaved code and data, at the cost of expensive encrypted memory paging mechanisms and serious performance overheads~\cite{pires_scbr:2016,brenner_securekeeper:_2016} when this limit is crossed.
Moreover, executing a fully-functional JVM inside an SGX enclave would currently involve significant re-engineering efforts.

\textsc{DEFCon}~\cite{Migliavacca:2010:DHE} relies also on the JVM.
This event processing system focuses on security by enforcing constraints on event flows between processing units.
The event flow control is enforced using application-level virtualisation to separate processing units in a \emph{ad-hoc} JVM.

A few recent contributions tackle privacy-preserving data processing, particularly in a MapReduce scenario.
This is the case of Airavat~\cite{Roy:2010:ASP:1855711.1855731} and \textsc{Gupt}~\cite{Mohan:2012:GPP:2213836.2213876}.
These systems leverage differential-privacy techniques~\cite{dwork2006calibrating} and can face a different threat model than the one supported by SGX and hence by \SYS.
In particular, when deploying such systems on a public infrastructure, one needs to trust the cloud provider.
Our system greatly reduces the trust boundaries, and only requires trust of Intel{\textregistered} and their SGX implementation.

Some authors contest that public clouds may be secure enough some parts of an application.
They propose to split the jobs, running only the critical parts in private clouds.
A privacy-aware framework on hybrid clouds~\cite{xu2015framework} has been proposed to work on tagged data, at different granularity levels.
A MapReduce preprocessor splits data into private and public clouds according to their sensitivity.
Sedic~\cite{zhang2011sedic} does not offer the same tagging granularity, but proposes to automatically modify reducers to optimize the data transfers in a hybrid cloud.
These solutions require splitting application and data in two parts (sensitive and not) and impose higher latencies due to data transfers between two different clouds.
Yet, they cannot offer better security guarantees that the software stack itself offers, be it public or private.

MrCrypt~\cite{tetali2013mrcrypt} proposes using homomorphic encryption instead of trusted elements.
Through static code analysis, it pinpoints different homomorphic encryption schemes for every data column.
Still, some of the demonstrated benchmarks are ten times slower than the unecrypted execution.
\SYS{} avoids of complex encryption schemes, decrypts data entering enclaves and processes in plaintext.

The \textsc{Styx}~\cite{Stephen:2016:SSP:2987550.2987574} system uses partial homorphic encryption to allow for efficient stream processing in trusted cloud environments.
Interestingly, the authors of that system mention Intel{\textregistered} SGX as possible alternative to deploy stream processing systems on trusted hardware offered by untrusted/malicious cloud environments.
\SYS{} offers insights on the performances of exactly this approach.

To best of our knowledge, \SYS{} is the first lightweight and low-memory footprint stream processing framework that can fully execute within SGX enclaves.

\sloppy As we described before, \SYS{} is executing processes taking advantage of SGX enclaves inside Docker containers.
\textsc{SCONE}~\cite{pietzuch_scone:_nodate}, which is not yet openly available, is a recently introduced system that offers a secure container mechanism for Docker to leverage the SGX trusted execution support.
It proposes a generic technology to embed any C program to execute inside an SGX enclave.
Rather than generic programs, \SYS{} offers support to execute a lightweight \luavm{} inside an SGX enclave and securely execute chunks of \textsc{Lua} code inside it.
In our experiments, we execute this \luavm{} inside Docker containers.
 
\section{Conclusion}
\label{sec:conclusion}

Secure stream processing is becoming a major concern in the era of the Internet of Things and big data.
This paper introduces our design and evaluation of \SYS{}, an concise and efficient middleware framework to implement, deploy and evaluate secure stream processing pipelines for continuous data streams.
The framework is designed to exploit the SGX \emph{trusted execution environments} readily available in Intel{\textregistered}'s commodity processors, such as the latest SkyLake.
We implemented the prototype of \SYS{} in \textsc{Lua} and based its APIs on the reactive programming approach.
Our initial evaluation results based on real-world traces are encouraging, and pave the way for deployment of stream processing systems over sensitive data on untrusted public clouds.

We plan in our future work to further extend and thoroughly evaluate \SYS against other known approaches on secure stream processing, like \textsc{Styx}~\cite{Stephen:2016:SSP:2987550.2987574}, MrCrypt~\cite{tetali2013mrcrypt} or \textsc{DEFCon}~\cite{Migliavacca:2010:DHE}.
In particular, we plan to extend \SYS with full automation of container deployments, as well as enriching the framework with a library of standard stream processing operators and efficient yet secure native plugins, to ease the development of complex stream processing pipelines.

\section*{Acknowledgments}
The research leading to these results has received funding from the European Commission, Information and Communication Technologies, H2020-ICT-2015 under grant agreement number 690111 (SecureCloud project).
Rafael Pires is also sponsored by CNPq, National Counsel of Technological and Scientific Development, Brazil.

{
\bibliographystyle{ACM-Reference-Format}

}

\end{document}